\documentclass[10pt]{article} 
\usepackage[tbtags]{amsmath} 
\usepackage{amsopn,amsthm,amsfonts,amssymb} 
 
\let\t\theta 
 
\newtheorem{theorem}{Theorem} 
\newtheorem{lemma}[theorem]{Lemma} 
\newtheorem{proposition}[theorem]{Proposition}

\newtheorem{corollary}[theorem]{Corollary} 
\newtheorem{identity}[theorem]{Identity} 
 
\newtheorem{ddefinition}[theorem]{Definition} 
 
\theoremstyle{remark} 
\newtheorem{remark}[theorem]{Remark}

\newcommand{\reff}[1]{(\ref{#1})} 
 
\newcommand{\mc}[1]{\ensuremath{\mathcal{#1}}} 

\newcommand{\sm}[1]{\ensuremath{m_{(#1)}}}

\newcommand{\mtext}[1]{\ensuremath{ \quad \mbox{#1}\quad } }

\begin{document} 
 
\title{Jack superpolynomials, 
superpartition ordering and determinantal formulas} 
 
\author{Patrick Desrosiers\thanks{pdesrosi@phy.ulaval.ca} \cr 
\emph{D\'epartement de Physique},\cr 
Universit\'e Laval, \cr 
Qu\'ebec, Canada, G1K 7P4. 
\and 
Luc Lapointe\thanks{lapointe@math.mcgill.ca }\cr 
\emph{Department of Mathematics and Statistics},\cr 
McGill University,\cr 
Montr\'eal, Qu\'ebec H3A 2K6, Canada. 
\and 
Pierre Mathieu\thanks{pmathieu@phy.ulaval.ca} \cr 
\emph{D\'epartement de Physique},\cr 
Universit\'e Laval, \cr 
Qu\'ebec, Canada, G1K 7P4. 
} 
 
\date{April 2001} 
 
\maketitle 
 
\begin{abstract}

We call superpartitions the indices of the eigenfunctions 
of the supersymmetric extension of the trigonometric Calogero-Moser-Sutherland 
model. We obtain an ordering 
on superpartitions from the explicit action of 
the model's Hamiltonian on monomial superfunctions. 
This allows to define Jack superpolynomials 
as the unique eigenfunctions of the model that decompose triangularly,
with respect to this ordering,  
on the basis of monomial superfunctions.  This further 
leads to a simple and explicit 
determinantal expression for the Jack superpolynomials.

\end{abstract} 
 
\newpage 
 

\section{Introduction} 
 
Orthogonal polynomials that are eigenfunctions of Hamiltonians 
of physical interest certainly deserve consideration. Among this class 
of orthogonal polynomials are the eigenfunctions 
of integrable quantum many-body 
systems of the Calogero-Moser-Sutherland (CMS) type 
\cite{Calogero:1969ie, Moser1974, Sutherland:1971ic,Olshanetsky:1981dk}, 
such as the Jack polynomials  \cite{Jack1970, Stanley1988},
the eigenfunctions of the trigonometric CMS model. 
 
In \cite{DLM}, we launched the study of a generalized version of Jack polynomials
that are eigenfunctions of
the supersymmetric extension of the 
trigonometric CMS model  \cite{SriramShastry:1993cz} (stCMS model). 
In addition to the bosonic variables 
$x_i$ ($i=1,\cdots, N$, where $N$ is the number of particles), 
this model contains fermionic degrees of freedom described by the 
variables $\theta_i$.  Therefore, the generalized version of Jack polynomials
incorporates  
Grassmannian (i.e., 
fermionic) variables.

The usual Jack polynomials are not uniquely defined as being 
eigenfunctions of the trigonometric CMS model.  
An extra ingredient is required to 
characterize them:  they need to decompose triangularly
on the basis of 
monomial functions. Such a triangular decomposition entails an 
ordering. In this case, it is the standard  dominance ordering 
\cite{Stanley1988, Macdonald1995}. 
 
A similar situation holds for the superanalogues: in order to 
characterize precisely the Jack superpolynomials, it is 
necessary to require a triangular decomposition  in terms 
of some sort of monomial superfunctions.  The first step in 
this case thus amounts to finding a fermionic extension of the 
monomial functions. 
Fortunately, there is a rather natural way of defining the symmetric monomial 
superfunctions once the general symmetry properties of the stCMS 
eigenfunctions are 
  understood. These monomial superfunctions are indexed by 
superpartitions (see \cite{DLM} and sect. 2), which are, roughly,  
ordered pairs of partitions (with one of the partitions  required to have 
distinct parts). The second step 
then reduces to the formulation of a  
proper ordering among superpartitions allowing to characterize the 
triangular decomposition of the Jack superpolynomials.   
Relying on the usual dominance ordering on standard partitions, it 
proved possible to find such an ordering \cite{DLM}.

Although this dominance ordering is technically 
sufficient to calculate the Jack superpolynomials, it does  
not provide the most precise characterization 
of the Jack superpolynomials. Indeed, it does not rule out monomials 
that do not actually appear in the decomposition of a given Jack 
superpolynomial. 
 
The aim of the present paper is to improve this situation 
by introducing a weaker
ordering (i.e., in the sense that fewer elements are comparable) on superpartitions. 
The basic clue to the discovery of the ordering lies in the following 
observation. If a triangular decomposition exists for the Jack 
superpolynomials in terms of the symmetric monomial 
superfunctions, the action of the Hamiltonian itself on the basis 
of  
  symmetric 
monomial superfunctions is likely to be triangular (and the converse 
is also true). The key point then is that there are usually 
fewer (and certainly never more) terms in the expansion of the 
action of the Hamiltonian  on a given symmetric supermonomial (in terms
of symmetric superpolynomials) than 
there are in the expansion of this superpolynomial (again in terms of
symmetric supermonomials).\footnote{In the terminology of 
this paper, the terms that can 
appear in the action of the Hamiltonian on the supermonomial basis 
are obtained from a single rearrangement.  In the Jack 
superpolynomial decomposition, there are in addition terms 
associated to superpartitions generated from multiple 
rearrangements.} Therefore, it is simpler to unravel the ordering 
at work by considering the action of the Hamiltonian on 
supermonomials. This is how the ordering presented here was 
found. 
 
The analysis of the action of the Hamiltonian on the supermonomial 
basis had an important offshoot: all the expansion coefficients 
could be calculated explicitly. This provided a rather nontrivial 
generalization of the results of \cite{Sogo}, which are recovered here 
as a special case.

As already stated, the triangular nature of the action of the Hamiltonian on 
the supermonomial basis implies a triangular decomposition of the Jack 
superpolynomials. This, combined with the fact that the superpartitions 
entering in the decomposition of a given Jack superpolynomial on the monomial 
superfunction basis 
are asssociated to distinct eigenvalues, 
readily implies that the Jack superpolynomials 
can be written as determinants. 
The entries of the determinants 
being the known eigenvalues and the known expansion coefficients of the 
action of the Hamiltonian on the 
  supermonomial basis, one thus ends up with a remarkably simple and 
rather explicit expression for the Jack superpolynomials that 
generalizes the one obtained in \cite{LapointeLascouxMorse} in the 
non-supersymmetric case. 
 
The article is organized as follows. In section 2, we briefly review the 
stCMS model as well as some results of \cite{DLM} that  are 
required for the present 
article, namely superpartitions, the symmetric 
supermonomials and the Jack 
superpolynomials. In section 3  we present our main result: the explicit 
  action of the 
Hamiltonian on the 
  supermonomial basis. The resulting monomials are characterized by 
partitions that can be obtained from the original one via the  action 
  of a two-particle lowering operator 
$R_{ij}^{(\ell)}$ (the labels $i$ and $j$ being those of the 
particles on which the operator acts - with `intensity' specified by 
$\ell$). The new ordering is also formulated directly in terms of 
this lowering operator. Section 4 is devoted to computing 
explicitly the action of the stCMS Hamiltonian on supermonomials 
and thereby proving the result stated in section 3. In the last 
section, we reformulate the definition of the Jack
superpolynomials in terms of this new ordering and present the determinantal  formula. 

We would like to stress that all our results provide  genuine extensions 
of non-supersymmetric
ones, the latter being associated to the
zero-fermion sector. In this sector, a superpartition reduces to a partition and a
monomial superfunction reduces to an ordinary monomial symmetric 
function. In this regard, the detailed proof of the action of the stCMS
Hamiltonian on the 
  supermonomial basis includes an explicit proof of the
main result of \cite{Sogo}.

\section{The supersymmetric 
CMS model and 
Jack superpolynomials} 
 
\subsection{The Hamiltonian} 
 
The Hamiltonian of the stCMS model reads 
\cite{SriramShastry:1993cz}: 
  \begin{equation} 
\mc{H}=-\frac{1}{2}\sum_{i=1}^{N} \partial_{x_i}^2+\left(\frac{\pi}{L}\right)^2\sum_{ 
i<j}\frac{\beta(\beta-1+\theta_{ij} \theta^{\dagger}_{ij})}{\sin^2(\pi 
x_{ij}/L)}-\left(\frac{\pi 
\beta}{L}\right)^2\frac{N(N^2-1)}{6}, 
\end{equation} 
where   $x_{ij}=x_i-x_j$,   $\theta_{ij} = \theta_{i} - \theta_{j}$ 
and  $\theta_{ij}^{\dagger} = \partial_{\theta_{i}} - \partial_{\theta_{j}}$. 
The  $\theta_{i}$'s, with  $ i=1,\cdots,N 
$, are the anticommuting variables that, added to the standard variables 
$ x_i $,  make the model  supersymmetric.  That is 
to say, for $i,j \in 
 \{1,\dots,N \}$, we have 
\begin{equation} 
[x_i,x_j] = 0 \, , \qquad \{ \theta_i,\theta_j \}=\theta_i\theta_j+\theta_j \theta_i=0 \, , \qquad 
[x_i,\theta_j]=0 \, . 
\end{equation} 
The 
supersymmetric invariance is enforced in the very construction of 
this Hamiltonian, 
i.e., in defining it 
  as the anticommutator of  two 
supersymmetric charges  
  $ Q $  and  $ Q^{\dagger} $ : 
  \begin{equation} 
     \mc{H}=\frac{1}{2} \{Q, Q^{\dagger} \} 
  \qquad {\rm {with}}\qquad 
Q=\sum_j \partial_{\theta_j}(\partial_{x_j}-i\Phi_j(x)), \qquad Q^{\dagger}=\sum_j 
\theta_j(\partial_{x_j}+i\Phi_j(x)) 
\end{equation} 
where 
\begin{equation} 
     \Phi_j(x)= \frac{\pi \beta}{ L} 
\sum_{k\not= j} \cot \left({\pi \frac{x_{ij}}{L}}\right)\, .
\end{equation} 
The supersymmetric charges are fixed by the
requirement that they be nilpotent and that, in the
absence of fermionic variables
(that is, by setting  $ \theta_{i}=\partial_{\theta_{i}}=0$), the 
Hamiltonian reduce to the tCMS Hamiltonian.

An observation that proves to be crucial in \cite{DLM} is that 
the term encapsulating the dependency upon the fermionic variables is a 
fermionic-exchange operator \cite{SriramShastry:1993cz}: 
\begin{equation} 
\kappa_{ij}\equiv 1-\theta_{ij}\theta^{\dagger}_{ij}= 
1-(\theta_{i}-\theta_j)(\partial_{\theta_i}-\partial_{\theta_j}). 
\end{equation} 
This means that its action on any monomial function 
  $ f(\theta_i,\theta_j) $ is such that 
\begin{equation} 
\kappa_{ij}\, 
f(\theta_i,\theta_j)= 
f(\theta_j,\theta_i)\, \kappa_{ij}. 
\end{equation} 
This allowed the powerful exchange-operator formalism 
\cite{Polychronakos:1992zk} to be used
to study the properties of the model in \cite{DLM}.  Note that when we 
refer to exchange-operators ($\kappa_{ij},K_{ij},\mc{K}_{ij}$ in this article)  
we understand operators that satisfy relations 
of the type: 
 \begin{equation} \label{proprioK} 
    \kappa_{ij}=\kappa_{ji} \mtext{,} 
    \kappa_{ij}\kappa_{jk}=\kappa_{ik}\kappa_{ij}= 
 \kappa_{jk}\kappa_{ki}\mtext{,} 
    \kappa_{ij}^2=1. \end{equation} 
 
It proves convenient to remove  the contribution of the 
ground-state wave function, 
\begin{equation} 
\psi_0(x) =\Delta^\beta(x) \equiv\prod_{j<k}\sin^\beta\left(\frac{\pi 
x_{jk}}{L}\right) \, ,
\end{equation} 
from the Hamiltonian. The transformed Hamiltonian (which is  still 
supersymmetric) 
becomes then 
\begin{equation} 
\bar{\mc{H}}\equiv 
  \frac{1}{2} \left(\frac{L}{\pi}\right)^2 \Delta^{-\beta}\mc{H}\Delta^{\beta} 
\, .
\end{equation} 
When written in terms of the new bosonic variables 
  $ z_j=e^{2\pi i x_j/L} $, it finally reads 
\begin{equation} \label{shjack} 
\bar{\mc{H}}= \sum_i (z_i \partial_i)^2+\beta \sum_{i<j}\frac{ z^{ij}} 
{z_{ij}}(z_i \partial_i-z_j\partial_j)-2\beta\sum_{i<j}\frac{z_i 
z_j}{z_{ij}^2}(1-\kappa_{ij}) \, . 
\end{equation} 
where $\partial_i= \partial_{z_i}$ and  $  z^{ij}= 
z_i+z_j$. The Jack superpolynomials will be eigenfunctions of 
this Hamiltonian.

\subsection{Superpartitions and the monomial symmetric superfunctions}

As explained in \cite{DLM}, since $\bar{\mc{H}}$ leaves invariant the space 
of polynomials of a given degree in $z$ and a given degree in $\theta$,  
we can look for eigenfunctions    
of the  form: 
\begin{equation} 
\mc{A}^{(m)} (z,\theta;\beta)=\sum_{1 \leq i_1<i_2< \ldots <i_m \leq 
N}\theta_{i_1}\cdots\theta_{i_m} A^{(i_1\ldots i_m)}(z;\beta)\, , 
\qquad m=0,1,2,3,\dots \, , 
\end{equation} 
where $ A^{(i_1\ldots i_m)}$  is a homogeneous polynomial in  $ z$.   
Due to the presence of  $ m $  fermionic variables in its  expansion, 
  $ \mc{A}^{(m)} $  is said to belong to the  $ m$-fermion sector. 
 
The solutions 
  $ \mc{A}^{(m)} $,  being symmetric 
superpolynomials, 
must be invariant under the combined action of  
$ \kappa_{ij} $  and  $ K_{ij} $, the exchange operator 
acting on the  $ z_i $  variables: 
\begin{equation} 
         K_{ij}f(z_i, z_j)=f(z_j, z_i,)K_{ij} \, . 
\end{equation} 
In other words,  $ \mc{A}^{(m)} $ 
must be invariant under the action of 
  $ \mc{K}_{ij} $ , where 
\begin{equation} 
\mc{K}_{ij}\equiv \kappa_{ij}K_{ij}. 
\end{equation} 
Given that the  $ \theta $  products are antisymmetric, the functions 
  $  A^{(i_1\ldots i_m)} $  must satisfy 
\begin{equation} 
\begin{array}{l} 
K_{ij}A^{(i_1\ldots i_m)} (z;\beta)=- 
{A}^{(i_1\ldots i_m)} (z;\beta) \quad \forall \quad i\mbox{ and }j \in 
\{i_1\ldots i_m\}\, ,\cr 
  K_{ij} {A}^{(i_1\ldots i_m)} (z;\beta)=\phantom{-} 
{A}^{(i_1\ldots i_m)} (z;\beta) \quad \forall \quad i\mbox{ and }j 
\not\in \{i_1\ldots i_m\}\, . 
\end{array} 
\end{equation} 
i.e.,  $  {A}^{(i_1\ldots i_m)} $  must be completely antisymmetric in 
the variables 
  $ \{ z_{i_1},\ldots, z_{i_m}\} $,  and totally symmetric in the remaining 
variables 
  $ z/\{z_{i_1},\ldots, z_{i_m}\} $.

Superpartitions (to be denoted by capital Greek letters) 
provide the proper labelling of symmetric superpolynomials 
\cite{DLM}. A superpartition in the  $ m $-fermion sector is a 
sequence of integers that generates two standard partitions 
separated by a semicolon: \begin{equation} 
\Lambda=(\Lambda_1,\ldots,\Lambda_m;\Lambda_{m+1},\ldots,\Lambda_{N})= 
(\lambda^a ; \lambda^s), \end{equation} the first one being 
associated to an antisymmetric function, meaning that its parts are all 
distinct: 
\begin{equation} 
\lambda^a=(\Lambda_1,\ldots,\Lambda_m), \qquad 
\Lambda_i>\Lambda_{i+1}\geq 0 \,,\quad \  i=1, \ldots m-1, 
\end{equation} 
and the second one, to a symmetric function: 
\begin{equation} 
\lambda^s= (\Lambda_{m+1},\ldots,\Lambda_{N}),  \qquad 
\Lambda_i \ge \Lambda_{i+1}\geq 0 \, , \quad  i=m+1,\dots,N-1 \, , 
\end{equation} 
with equal parts then being allowed. 
In the zero-fermion sector, the 
semicolon is omitted and 
  $ \Lambda $  reduces to  $ \lambda^s $. We often write the degree of a 
superpartition as  $ n =|\Lambda|=\sum_{i=1}^{N}\Lambda_i$. 
Note finally that to any  superpartition  $ \Lambda $,  we associate a unique 
standard partition  $ \Lambda^* $  obtained by rearranging the parts of the 
superpartition in decreasing order.  For instance, the rearrangement 
of $\Lambda=(4,2,1;5,3,3,1)$ is the partition $\Lambda^* 
=(5,4,3,3,2,1,1)$.

The monomial symmetric superpolynomials have been introduced in 
\cite{DLM} as a natural basis of the ring of symmetric 
superfunctions (we will also refer to this basis as the supermonomial basis): 
\begin{equation} 
m_{\Lambda}(z,\theta)=\sm{\Lambda_1,\ldots, 
\Lambda_m;\Lambda_{m+1},\ldots,\Lambda_{N}}(z,\theta)={\sum_{\sigma\in 
S_{N}}}' \theta^{\sigma(1, \ldots, m)}z^{\sigma(\Lambda)}, \end{equation} 
where the prime 
indicates that the  summation is 
restricted to distinct terms, and where 
\begin{equation} 
z^{\sigma(\Lambda)}=z_1^{\Lambda_{\sigma(1)}} \cdots z_m^{\Lambda_{\sigma(m)}} 
z_{m+1}^{\Lambda_{\sigma(m+1)}} \cdots 
z_{N}^{\Lambda_{\sigma(N)}}  \quad {\rm{and}} \quad 
\theta^{\sigma(1, \ldots, m)} = \theta_{\sigma(1)} \cdots 
\theta_{\sigma(m)} \, . 
\end{equation} 
Given a reduced decomposition $\sigma = \sigma_{i_1} \cdots \sigma_{i_n}$ 
of an element $\sigma$ of the symmetric group $S_N$, let 
$\mc{K}_{\sigma}$ stand for $\mc{K}_{i_1,i_1+1} \cdots \mc{K}_{i_n,i_n+1}$. 
In this notation, we can rewrite the monomial superfunction $m_{\Lambda}$ 
as 
\begin{equation} 
m_{\Lambda}= \frac{1}{f_{\Lambda}} \sum_{\sigma \in S_N} \mc{K}_{\sigma} 
\left( \theta_1 \cdots \theta_m z^{\Lambda}\right) \, , 
\end{equation} 
where the normalization constant $f_{\Lambda}$ is 
\begin{equation} 
f_{\Lambda} =f_{\lambda^s}= n_{\lambda^s}(0)!\, n_{\lambda^s}(1)! \,  
n_{\lambda^s}(2) ! \cdots \, , 
\end{equation} 
with  $n_{\lambda^s}(i)$ the number of $i$'s in $\lambda^s$, the symmetric part of 
$\Lambda=(\lambda^a;\lambda^s)$.

\subsection{Jack superpolynomials} \label{sectdef}

The {\it Jack superpolynomials} can now 
be defined. 
\begin{ddefinition} 
The Jack superpolynomials are the unique functions 
satisfying the following two properties \cite{DLM}: 
\begin{equation} \label{sjackensm} 
\begin{split} 
(i): & \quad \bar{\mc{H}}\, \mc{J}_\Lambda (z,\theta;\beta)  = 
\varepsilon_\Lambda\mc{J}_{\Lambda}(z,\theta;\beta)\, , \\ 
(ii): & \quad 
\mc{J}_\Lambda (z,\theta;\beta)=m_{\Lambda} (z,\theta)+\sum_{\Omega;\, \Omega^* 
< \Lambda^*}c_{\Lambda,\Omega}(\beta) m_{\Omega}(z,\theta)\, . 
\end{split} 
\end{equation} 
\label{defjack} 
\end{ddefinition} 
\noindent In the definition, the 
ordering on 
partitions is the usual dominance ordering, 
\begin{equation} 
      \lambda \geq \omega \mtext{iff} 
     \lambda_1+\lambda_2+\dots 
+ 
\lambda_i \geq \omega_1 + \omega_2 +\dots +\omega_i \, , \quad \forall i \, , 
\end{equation} 
and the eigenvalue $\varepsilon_{\Lambda}$, which 
turns out to be  independent of 
the fermionic sector, is given by: 
\begin{equation}\label{valeurp} 
\varepsilon_\Lambda= \varepsilon_{\Lambda^*}=\sum_j 
[{\Lambda^*_j}^2+\beta(N+1-2j)\Lambda^*_j]\,. 
\end{equation}  
where $\Lambda^*_i\equiv (\Lambda^*)_i$. The uniqueness property, rather obvious from a
computational point of view, is established in sect. 5. Many explicit examples of Jack
superpolynomials  can be found in
\cite{DLM}. 
 
As stressed in \cite{DLM}, the mere existence and uniqueness of functions 
satisfying the two conditions of Definition~\ref{defjack} 
is remarkable.  Later in this 
article, 
we will show that this definition does in fact provide a genuine 
characterization 
of the Jack superpolynomials. 
However, it was also 
observed that the dominance ordering of the  
partitions $\Lambda^*$ is not precise enough in the sense that some 
monomials  $ m_{\Omega} $  with  $ \Omega^* < \Lambda^* $  do 
not appear in the sum. 
  A partial ordering 
among superpartitions that fixes this problem will be 
presented in the following section.

\section{Action on the monomial basis and 
ordering of 
superpartitions}

In order to motivate the dominance ordering presented at the 
end of this section, we first display the explicit action of  $ 
\bar{\mc{H}} $  on the monomial superpolynomials. This, in turn, 
will lead to a simple expression for the Jack 
superpolynomials in the form of determinants (cf. sect. 5).

As in the previous section, let  $ \Lambda= 
(\lambda^a;\lambda^s)=(\Lambda_1,\dots, 
\Lambda_m;\Lambda_{m+1},\dots,\Lambda_{N}) $ 
be a  superpartition in the 
  $ m $-fermion sector. 
 
In order to  characterize the monomials that are generated by the action 
of   $ 
\bar{\mc{H}} 
$ on 
   $ m_{\Lambda} $,  we introduce the 
operator\footnote{The non-supersymmetric version of this operator was 
introduced in 
\cite{Macdonald1995} in the case $\ell=1$ (more precisely, it is the raising 
version of the operator that is used), and in 
\cite{LapointeLascouxMorse} in the general case.} 
  $ R_{ij}^{(\ell)} $  whose action,  for  $ i<j $  and  $ \ell \geq 0 $, 
is given by: 
\begin{equation} \label{rij} 
R_{ij}^{(\ell)}(\Lambda_1,\dots,\Lambda_i,\dots,\Lambda_j,\dots,\Lambda_{N}) 
=\left\{ 
\begin{array}{ll} 
(\Lambda_1,\dots,\Lambda_i-\ell,\dots,\Lambda_j+\ell,\dots,\Lambda_{N}) & 
{\rm{if }} \, \, \Lambda_i > \Lambda_j\,, \cr 
(\Lambda_1,\dots,\Lambda_i+\ell,\dots,\Lambda_j-\ell,\dots,\Lambda_{N}) & 
{\rm{if }} \, \, \Lambda_j > \Lambda_i\,. 
\end{array}\right. 
\end{equation} 
This action of  $ R_{ij}^{(\ell)} $  is non-zero only in the following cases: 
\begin{equation} 
\begin{array}{rl} 
{\rm{I}}: & \quad i,j \in \{1,\dots,m \}\,  \quad {\rm{and}} \quad 
\lfloor \frac{\Lambda_i-\Lambda_j-1}{2}\rfloor \geq \ell \,,\cr 
{\rm{II}}: & \quad i \in \{1,\dots,m\} \, , j \in \{m+1,\dots,N \}\, 
\quad {\rm{and}} \quad 
|\Lambda_i-\Lambda_j|-1 \geq   \ell\,,\cr 
{\rm{III}}: & \quad i,j \in \{m+1,\dots,N \}\,  \quad {\rm{and}} \quad 
\lfloor \frac{\Lambda_i-\Lambda_j}{2}\rfloor \geq \ell \,.\cr 
\end{array} 
\label{type} 
\end{equation} 
where $\lfloor x \rfloor$ stands for the largest integer smaller or equal to 
$x$. Otherwise, it is 
understood that 
$ R_{ij}^{(\ell)} $  annihilates the superpartition 
  $ \Lambda $,  i.e.,  $ R_{ij}^{(\ell)}\, \Lambda=\emptyset $. In the following, 
we will say that a  pair $(i,j)$ is of type I if $ i,j \in \{1,\dots,m \}$, of  type II if 
$i \in \{1,\dots,m\}$  and $j \in \{m+1,\dots,N \}$, and of type III if $i,j \in \{m+1,\dots,N
\}$.

In the $m$-fermion sector, given a sequence $\gamma=(\gamma_1,\dots,\gamma_m; 
\gamma_{m+1},\dots,\gamma_N)$, we will denote by $\overline{\gamma}$ 
the superpartition whose antisymmetric part is the 
rearrangement of $(\gamma_1,\dots,\gamma_m)$ and whose symmetric part is 
the rearrangement of $(\gamma_{m+1},\dots,\gamma_N)$. 
For example, we have 
\begin{equation}
\overline{(1,3,2;2,3,1,2)}
=(3,2,1;3,2,2,1).
\end{equation}  
Also, $\sigma_{\gamma}$ will stand for the element 
of $S_N$ 
that sends $\gamma$ to $\overline{\gamma}$, that is $\sigma_{\gamma} \gamma= 
\overline{\gamma}$.  Note that we can always choose  
$\sigma_{\gamma}$ such that $\sigma_{\gamma}=\sigma_{\gamma}^a\sigma_{\gamma}^s$, 
with $\sigma_{\gamma}^a$ and $\sigma_{\gamma}^s$ permutations of $\{1,\dots,m\}$ 
and $\{m+1,\dots,N \}$ respectively.

We now state the explicit 
action of the stCMS Hamiltonian on monomials. The 
proof will be presented in the following section. 
\begin{theorem} 
  $ \bar {\cal{H}} $  acts on 
  $ m_{\Lambda}(z,\theta) $  as follows: 
\begin{equation} 
\bar {\cal{H}} \, m_{\Lambda}(z,\theta)= \varepsilon_{\Lambda} 
m_{\Lambda}(z,\theta)+\sum_{\Omega\neq \Lambda} v_{\Lambda \Omega}(\beta) 
\, m_{\Omega}(z,\theta) 
\label{expa} 
\end{equation} 
where  $ \varepsilon_{\Lambda} $ is given in \reff{valeurp} and  $ v_{\Lambda
\Omega}(\beta)
$  is non-zero only if  $ \Omega =  (\omega^a;\omega^s)= 
\overline{R_{ij}^{(\ell)} \Lambda} $ , for a given 
  $ R_{ij}^{(\ell)} $ with $\ell>0$. 
In this case, the 
coefficient  $ v_{\Lambda \Omega}(\beta) $ 
reads: 
\begin{equation}\label{vexpli} 
v_{\Lambda \Omega}(\beta) = \left\{ 
\begin{array}{ll} 
2\beta \, {{\rm{sgn}}}(\sigma_{R_{ij}^{(\ell)} \Lambda}^a)
\bigl(\Lambda_i-\Lambda_j-\delta\ell  \bigr) 
n(\Lambda_i-\ell,\Lambda_j+\ell) &  {\rm{if }}\, \, \Lambda_i > \Lambda_j \cr 
2\beta  \, {{\rm{sgn}}}(\sigma_{R_{ij}^{(\ell)} \Lambda}^a)  
\bigl(\Lambda_j-\Lambda_i-\delta\ell  \bigr) 
n(\Lambda_i+\ell,\Lambda_j-\ell) &  {\rm{if }} \, \,\Lambda_j > \Lambda_i 
\end{array}\right. 
\end{equation} 
where  ${{\rm{sgn}}}(\sigma_{R_{ij}^{(\ell)} \Lambda}^a) $  
 stands for the  sign of the permutation $\sigma_{R_{ij}^{(\ell)} \Lambda}^a$.  
 The parameter   $ 
\delta $ is 
equal to 
  $ 2,1 $  or  $ 0 $  if  the pair  $ (i,j) $  is of type I, II or III in 
(\ref{type})  respectively. Finally   $ n(a,b) $  is a symmetry 
factor given by: 
\begin{equation}\label{symfa} 
n(a,b)=\left\{ 
\begin{array}{ll} 
1 & i,j \, \, {\rm{of~type~I}} \cr 
n_{\omega^s}(b)  & i,j \,\, {\rm{of~type~II}} \cr 
n_{\omega^s}(a)n_{\omega^s}(b) & i,j \,\, {\rm{of~type~III~and}} \, \, 
a \neq b \cr 
\frac{1}{2} n_{\omega^s}(a)\left(n_{\omega^s}(a)-1\right) 
& i,j \,\, {\rm{of~type~III~and}} \, \, 
a= b \cr 
\end{array} 
\right. \, . 
  \end{equation} 
\end{theorem} 
 
Since in \reff{vexpli}, $1\leq \ell \leq \max\{\Lambda_i,\Lambda_j \}$, 
the symmetry factor $n(a,b)$ of the non-diagonal coefficients is never invoked with
arguments
$a$  or $b$ equal to zero. This has the important consequence that 
the coefficients $v_{\Lambda \Omega}(\beta)$ in \reff{expa} do not 
depend on the number $N$ of variables, as long as $N$ is large 
enough so that $m_{\Omega}(z,\theta)\neq 0$. Indeed, with 
$\Omega=(\omega^{a};\omega^s)$, the number of variables only 
influences the number of parts equal to zero in the partition 
$\omega^s$, which as we said  does not influence the value of the symmetry 
factor $n(a,b)$. 
 
\begin{remark} 
\label{remark} If  $ \Omega= 
\overline{R_{ij}^{(\ell)} \Lambda}= \overline{R_{i'j'}^{(\ell')} \Lambda }$, 
we must have 
   $ \ell=\ell' $,   $ \Lambda_i=\Lambda_{i'} $  and  $ 
\Lambda_j=\Lambda_{j'} $. 
  The  choice of 
  $ i,j $  or $ i',j' $  is thus irrelevant in the computation of  $ v_{\Lambda 
\Omega}(\beta) $. 
\end{remark}

We now look at an example.  For $N \geq5$, we have 
\begin{equation} 
\begin{array}{rl} 
\bar {\cal{H}}\,m_{(2,1;4,2)} =& \varepsilon_{(2,1;4,2)}\, m_{(2,1;4,2)} 
+2\beta \, m_{(3,1;3,2)} -4\beta \, m_{(3,2;2,2)}+4\beta 
\, m_{(2,1;3,3)} \cr 
&+8\beta \, m_{(2,1;3,2,1)}+24\beta \, m_{(2,1;2,2,2)}+ 
4\beta \, m_{(2,1;4,1,1)} \, . 
\end{array} 
\end{equation} 
The coefficient  $ -4\beta $  in front of  $ m_{(3,2;2,2)} $  is 
explained first by 
identifying the proper rearrangement that links $(3,2;2,2)$ to 
$(2,1;4,2)$, namely 
\begin{equation} 
(3,2;2,2)=\overline{R_{2,3}^{(2)}\, (2,1;4,2)}=\overline{(2,3;2,2)} 
\end{equation} 
(which is a type-II case) and then by evaluating the corresponding 
coefficient $v_{\Lambda, \Omega}$ from Theorem 2: 
\begin{equation} 
  2\beta \, {{\rm{sgn}}}(\sigma^a_{(2,3;2,2)})\bigl(4-1- 2  \bigr) 
n(3,2)=-4\beta \, , 
\end{equation}  since  $ n(3,2)=n_{(2,2)}(2) =2$   
(there are two 2's in  $ (2,2) $ ) and 
the sign 
of the permutation that sends  $ (2,3) $  to  $ (3,2) $  is  $ -1 $.

Knowing the action of  $ \bar{{\cal{H}}} $  on the monomial basis allows us 
to define a partial ordering on superpartitions. 
\begin{ddefinition}  We say that 
$\Lambda \geq^{s} \Omega$ iff 
$\Omega = 
\overline{R_{i_k,j_k}^{(\ell_k)} \dots \overline{R_{i_1,j_1}^{(\ell_1)} 
\Lambda}}$, 
for a given sequence of operators 
  $ R_{i_1,j_1}^{(\ell_1)},\dots,R_{i_k,j_k}^{(\ell_k)} $ (the equality 
occurs in the case of the null sequence).
\footnote{The dominance ordering on partitions
is a special case of this ordering on superpartitions. 
We draw attention to the technical importance of formulating 
an ordering using the
lowering  operator
$R_{ij}^{(\ell)}$, especially in the case of the usual dominance 
ordering (as opposed to its formulation in
terms of inequalities).}  
\end{ddefinition}

\begin{proposition} 
The relation $\geq^s$ provides a partial ordering on 
superpartitions. 
\end{proposition} 
\noindent {\it Proof:} \quad 
To show that $\geq^s$ is a genuine ordering, we have to verify the following three
properties: \\
$~$ \qquad \qquad \qquad (1) reflexivity:  $\Lambda \geq^s \Lambda$; \\
$~$ \qquad \qquad \qquad  (2) transitivity: if 
$\Gamma \geq^s\Omega
 $ and  $\Omega \geq^s \Lambda $,  then $\Gamma  \geq^s\Lambda$; \\
$~$ \qquad \qquad \qquad  (3) antisymmetry:  
if $\Lambda \geq^s \Omega$ and $\Omega \geq^s \Lambda$, then $\Lambda=
\Omega$.

 Reflexivity follows from the consideration of a null sequence of lowering operators. The 
transitivity property is also immediate since two sequences can be glued together to form
a longer sequence.
In order to prove antisymmetry, we first make the following two 
simple observations 
from \reff{rij}: 
\begin{equation} \label{condordering} 
(i): \, \Lambda \geq^s \Omega \, \Longrightarrow \Lambda^* \geq \Omega^* 
\, , \qquad 
(ii): \, \Lambda \geq^s \Omega \quad {\rm{and}} \quad  \Lambda^*=\Omega^* \, \, 
\Longrightarrow \Lambda = \Omega \, , 
\end{equation} 
where as usual $\Lambda^*$ and $\Omega^*$ are the partitions associated to the 
superpartitions $\Lambda$ and $\Omega$ respectively.  Now, let 
$\Lambda \geq^s \Omega$ and $\Omega \geq^s \Lambda$.   Then, from the 
first observation, $\Lambda^* \geq \Omega^*$ 
and $\Omega^* \geq \Lambda^*$, i.e., $\Lambda^*=\Omega^*$.  Therefore, we have 
$\Lambda \geq^s \Omega$ and $\Lambda^* = \Omega^*$, which lead to 
$\Lambda=\Omega$ from the second observation.
\hfill
$\blacksquare$ 
 
Using this new partial ordering on superpartitions, we now have that the action 
\reff{expa} of $ \bar{\mc{H}} $ on $m_{\Lambda}(z,\theta)$ is triangular. 
\begin{corollary} \label{corotri} 
\begin{equation} \label{expatri} 
\bar {\cal{H}} \, m_{\Lambda}(z,\theta)= \varepsilon_{\Lambda} 
m_{\Lambda}(z,\theta)+\sum_{\Omega <^s \Lambda} v_{\Lambda \Omega}(\beta) 
\, m_{\Omega}(z,\theta) \, . 
\end{equation} 
\end{corollary} 
 
Notice that not all $\Omega$'s such that $\Omega <^s \Lambda$ 
appear in this expansion with a non-zero coefficient.
Only those that can be obtained from $\Lambda$ by 
the application of a {\it single}
lowering operator have non-zero coefficients.
 
\section{The action of  $ \bar{\mc{H}} $  on the monomial basis: a detailed 
proof}

We will first focus on the non-diagonal coefficients in the action of $ 
\bar{\mc{H}} $ on 
  $m_{\Lambda}(z,\theta)$, that is, on the coefficients 
$v_{\Lambda,\Omega}$ for $\Lambda\not=\Omega$, and wait until 
the end of the section before
computing the diagonal coefficient $\varepsilon_\Lambda$. 
We will show  that
the coefficients $v_{\Lambda,\Omega}$ are indeed given 
by the expression  displayed in
Theorem 2 for the  special case $N=2$, for the  pairs $(i,j)$ of types I, II 
and III. And, in
a second step, we will prove the  theorem in general. The reason why the
two-particle case  plays such a central role is that the operator 
$R_{ij}^{(\ell)}$ acts only on {\it two parts of a partition}, 
that is, on $N=2$ particles. 
 
\subsection{Non-diagonal coefficients: the case $N=2$}
 
 
We will use the following notation: 
\begin{equation} 
B_{ij}= \frac{z^{ij}}{z_{ij}} (z_i\partial_i-z_j\partial_j),\qquad 
F_{ij}= -2\frac{z_i z_j}{z_{ij}^2} (1-\kappa_{ij})  , 
\end{equation} 
  ($B$ for boson and 
$F$ for fermion). The part of the Hamiltonian acting non-diagonally on the 
monomial basis is 
$\beta(B+F)$, where 
\begin{equation} 
B= \sum_{1\leq i<j\leq N}B_{ij}, \qquad F=  \sum_{1\leq i<j\leq N}F_{ij}\, . 
\end{equation} 
In order to simplify the notation,  we will set  
$\Lambda_1=r,\, \Lambda_2=s$. 
 
\vskip0.2cm 
\noindent{\bf Case III}: 
\vskip0.2cm  In this case, we must have 
$m=0$.  This gives $\Lambda=(;r,s)=(r,s)$, and 
\begin{equation}\label{leka} 
m_{(r,s)}= \frac{1}{f_{(r,s)}}(1+\mc{K}_{12})\, z_1^r z_2^s =  
\frac{1}{f_{(r,s)}} ( z_1^rz_2^s+z_1^sz_2^r) \, . 
\end{equation} 
Clearly, the action of $F_{12}$ vanishes since there are no $\t$ terms. The 
action 
of $B_{12}$ is simply 
\begin{eqnarray} 
B_{12}m_{(r,s)} &=&  \frac{1}{f_{(r,s)}}\frac{z^{12}}{z_{12}} (r-s) (z_1z_2)^s 
(z_1^{r-s}-z_2^{r-s})\cr 
&=&  \frac{z^{12}}{z_{12}} (r-s) (z_1z_2)^s 
z_{12}(z_1^{r-s-1}+z_1^{r-s-2}z_2+\cdots 
+z_1z_2^{r-s-2} + z_2^{r-s-1})\cr 
&=&  (r-s) (z_1z_2)^s 
\big(z_1^{r-s}+2[z_1^{r-s-1}z_2+\cdots + z_1z_2^{r-s-1}] 
+z_2^{r-s}\big) \,  
\end{eqnarray} 
where we have set $f_{(r,s)}=1$, since the action vanishes in the case 
$r=s$. 
Therefore, by recombining to construct monomial symmetric functions, 
we get 
\begin{equation}\label{lafi} 
\beta B_{12}m_{(r,s)}= \beta B m_{(r,s)}= \beta(r-s) m_{(r,s)} + 
\sum_{\ell=1}^{\lfloor(r-s)/2 \rfloor} 2\beta(r-s) 
m_{(r-\ell,s+\ell)} 
\, . 
\end{equation} 
If we let $\Omega=(r-\ell,s+\ell)$, we have 
$\Omega=R_{12}^{(\ell)}\Lambda= \overline{R_{12}^{(\ell)} \Lambda}$.
We thus recover the right value $2\beta(\Lambda_1-\Lambda_2)=2\beta(r-s)$ 
for the coefficient $v_{\Lambda,\Omega}$, for $1 \leq \ell \leq 
\lfloor(\Lambda_1- 
\Lambda_2)/2 \rfloor$ 
(see \reff{vexpli} with $\delta=0$).

\vskip0.2cm 
\noindent{\bf Case II}: 
\vskip0.2cm 
We now have $m=1$. 
Consider first the action of $B+F$ on 
$m_{(r;s)}$, supposing that $r>s$.  Given 
\begin{equation}\label{lekaa} 
m_{(r;s)}= \frac{1}{f_{(r;s)}}(1+\mc{K}_{12})\, \theta_1 z_1^r z_2^s = 
\t_1z_1^rz_2^s+\t_2z_1^sz_2^r\, 
\end{equation} 
(since $f_{(r;s)}=1$), 
we get 
\begin{equation} 
B_{12}m_{(r;s)}= 
\frac{z^{12}}{z_{12}} (r-s) (z_1z_2)^s (\t_1z_1^{r-s}-\t_2 
z_2^{r-s})\end{equation} 
and 
\begin{eqnarray} 
F_{12}m_{(r;s)} & = & 
-2\frac{z_1z_2}{ z_{12}^2} (\t_1-\t_2) (z_1^r z_2^s -z_1^s z_2^r) \cr 
& = & 
-\frac{2}{z_{12}} (\t_1-\t_2)(z_1z_2)^s (z_1^{r-s}z_2+\cdots + 
z_1z_2^{r-s}) \, .
\end{eqnarray} 
In the sum $B_{12}+F_{12}$, we now concentrate on the term $\t_1$: 
\begin{equation} 
\left. (B_{12}+F_{12}) m_{(r;s)}\right|_{\t_1}= \frac{\t_1}{z_{12}}(z_1z_2)^s\{ 
(r-s)z^{12}z_1^{r-s}-2(z_1^{r-s}z_2+\cdots + 
z_1z_2^{r-s})\}\, . 
\end{equation} 
To proceed further, we use the following identity proved in the 
appendix. 
\begin{identity} 
\label{iden1} 
\begin{equation} 
\{ 
(r-s)z^{12}z_1^{r-s}-2(z_1^{r-s}z_2+\cdots + 
z_1z_2^{r-s})\}= z_{12}\{ (r-s)z_1^{r-s}+ \sum_{\ell=1}^{r-s-1} 
2(r-s-\ell)z_1^{r-s-\ell}z_2^\ell\}\, . 
\end{equation} 
\end{identity} 
 
Taking into account the $\t_2$ 
terms (which are recovered by symmetry from the $\t_1$ terms under 
$z_1 \leftrightarrow z_2$), we  have: 
\begin{equation}\label{lafia} 
  \beta(B_{12}+F_{12})m_{(r;s)}= \beta(B+F)m_{(r;s)}= \beta(r-s) m_{(r;s)}+ 
\sum_{\ell=1}^{r-s-1} 2\beta(r-s-\ell)  m_{(r-\ell;s+\ell)} \, . 
\end{equation} 
We thus get the right coefficient, namely 
$2\beta(\Lambda_1-\Lambda_2-\delta \ell)$, with  
$\delta=1$. Moreover, the upper limit on $\ell$  shows that 
$\Lambda_1-\Lambda_2 >\ell $, 
as it should. The derivation in the case $r<s$ is identical. 
 
\vskip0.2cm 
\noindent{\bf Case I}: 
\vskip0.2cm 
 
Since we now have $m=2$, we consider the action of $B+F$ on $m_{(r,s;0)}$. 
Given that, for  $r>s$,
\begin{equation}\label{lekaaa} 
  m_{(r,s;0)}=\frac{1}{f_{(r,s;0)}}(1+\mc{K}_{12})\,  
\theta_1 \theta_2 z_1^r z_2^s = \t_1\t_2(z_1^rz_2^s-z_1^s z_2^{r}) \, ,
\end{equation} 
a direct computation yields 
\begin{equation} 
B_{12}m_{(r,s;0)}= 
\t_1\t_2\frac{z^{12}}{z_{12}} (r-s) (z_1z_2)^s (z_1^{r-s}+z_2^{r-s}) \, ,
\end{equation} 
and since $(1-\kappa_{12})\t_1\t_2= 2\t_1\t_2$, 
\begin{eqnarray} 
F_{12}m_{(r,s;0)} &=& 
-4\t_1\t_2\frac{z_1z_2}{ z_{12}^2}  (z_1^r z_2^s -z_1^s z_2^r)\cr 
&=& 
-4\t_1\t_2\frac{z_1z_2}{ z_{12}^2}  (z_1 z_2)^s 
z_{12}(z_1^{r-s-1}+z_1^{r-s-2}z_2+\cdots+ z_2^{r-s-1})\cr 
&=& 
-4\t_1\t_2\frac{1}{z_{12}}  (z_1 z_2)^s 
(z_1^{r-s}z_2+z_1^{r-s-1}z_2^2+\cdots +z_1z_2^{r-s}) \, .
\end{eqnarray} 
Therefore, we obtain 
\begin{equation} 
(B_{12}+F_{12}) m_{(r,s;0)}= \frac{\t_1\t_2}{z_{12}}  (z_1 z_2)^s\{ 
(r-s)z^{12}(z_1^{r-s}+z_2^{r-s}) -4(z_1^{r-s}z_2+\cdots + z_1z_2^{r-s})\} \, .
\end{equation} 
 
To proceed, we need the following identity also 
proved in the appendix. 
\begin{identity} 
\label{iden2} 
\begin{eqnarray} 
  &&\{ 
(r-s)z^{12}(z_1^{r-s}+z_2^{r-s})-4(z_1^{r-s}z_2+\cdots + z_1z_2^{r-s})\} 
  \cr 
&&\qquad \quad = z_{12} \big\{ 
(r-s)(z_1^{r-s}-z_2^{r-s}) +\sum_{\ell=1}^{\lfloor(r-s-1)/2\rfloor} 
2(r-s-2\ell)(z_1^{r-s-\ell}z_2^\ell -z_1^\ell z_2^{r-s-\ell})\big\}\, . 
\end{eqnarray} 
\end{identity} 
 
Using the identity, we get 
\begin{equation}\label{lafiaa}
\beta(B_{12}+F_{12}) m_{(r,s;0)}=\beta(B+F) m_{(r,s;0)}=\beta(r-s)m_{(r,s;0)}+ 
\sum_{\ell=1}^{ \lfloor (r-s-1)/2\rfloor } 
2\beta(r-s-2\ell)m_{(r-\ell,s+\ell;0)}\, . 
\end{equation} 
{}From this expression, the value of the coefficient 
$v_{\Lambda,\Omega}$ is given immediately. It is indeed $2\beta(
\Lambda_1-\Lambda_2-\delta\ell)$, with 
$\delta=2$. The upper limit in the summation 
implies that 
$\lfloor(\Lambda_1-\Lambda_2-1)/2\rfloor\geq 
\ell$, as it should.


\subsection{Non-diagonal coefficients: the case $N>2$}
 
Let $\Lambda_{ij}$ denote the restriction of a superpartition to its $i^{th}$ 
and $j^{th}$ entries, that is, 
\begin{equation} 
\Lambda_{ij}= 
\begin{cases} 
(\Lambda_i,\Lambda_j;0) & {\text{if $i,j$ is of type I}} \\ 
(\Lambda_i;\Lambda_j) & {\text{if $i,j$ is of type II}} \\ 
(\Lambda_i,\Lambda_j) & {\text{if $i,j$ is of type III}} \\ 
\end{cases} \, . 
\end{equation}
Also, in accordance with \reff{type}, we define ${\cal{S}}^{\Lambda}_{ij}$, to be 
\begin{equation} 
{\cal{S}}^{\Lambda}_{ij} =  
\begin{cases} 
\lfloor \frac{\Lambda_i-\Lambda_j-1}{2}\rfloor & {\text{if $i,j$ of type I}} \\ 
|\Lambda_i-\Lambda_j| -1  & {\text{if $i,j$ of type II}} \\ 
\lfloor \frac{\Lambda_i-\Lambda_j}{2}\rfloor &  {\text{if $i,j$ of type III}}  
\end{cases} \, . 
\end{equation}  
 
For an arbitrary number of particules, we have thus 
\begin{equation} 
\label{bigeq} 
\begin{split} 
\beta(B+F)\, m_{\Lambda} & = 
\beta\sum_{1\leq i < j \leq N} (B_{ij}+F_{ij}) 
\frac{1}{f_{\Lambda}} \sum_{\sigma \in S_N} \mc{K}_{\sigma} 
\left( \theta_1 \cdots \theta_m z^{\Lambda}\right) \\ 
& = 
\frac{\beta}{f_{\Lambda}} \sum_{\sigma \in S_N} \mc{K}_{\sigma} \left( 
  \sum_{1\leq i < j \leq N} (B_{ij}+F_{ij}) 
  \, \theta_1 \cdots \theta_m z^{\Lambda}\right) \\ 
& = 
\frac{\beta}{f_{\Lambda}} \sum_{1\leq i < j \leq N} 
  \sum_{\sigma \in S_N} \mc{K}_{\sigma} \left( 
  (B_{ij}+F_{ij}) 
  \, \theta_1 \cdots \theta_m z^{\Lambda}\right) \\ 
& = 
\frac{1}{2}\frac{\beta}{f_{\Lambda}} \sum_{1\leq i < j \leq N} 
  \sum_{\sigma \in S_N} \mc{K}_{\sigma} \left( 
  (B_{ij}+F_{ij})(1+\mc{K}_{ij}) 
  \, \theta_1 \cdots \theta_m z^{\Lambda}\right) \\ 
& = \frac{1}{2}\frac{1}{f_{\Lambda}} \sum_{1\leq i < j \leq N} 
  \sum_{\ell=0}^{{\cal{S}}_{ij}^{\Lambda}} 
v_{\Lambda_{ij} \left(R_{ij}^{(\ell)}\Lambda \right)_{ij}} 
\sum_{\sigma \in S_N} \mc{K}_{\sigma} \left( \frac{1}{f_{\left(R_{ij}^{(\ell)} 
\Lambda\right)_{ij}}} (1+\mc{K}_{ij}) 
  \, \theta_1 \cdots \theta_m z^{R_{ij}^{(\ell)}\Lambda}\right) \\ 
& = \frac{1}{f_{\Lambda}} \sum_{1\leq i < j \leq N} 
  \sum_{\ell=0}^{{\cal{S}}_{ij}^{\Lambda}}  
\frac{v_{\Lambda_{ij} \left(R_{ij}^{(\ell)}\Lambda \right)_{ij}}} 
{f_{\left(R_{ij}^{(\ell)} \Lambda\right)_{ij}}} 
\sum_{\sigma \in S_N} \mc{K}_{\sigma} \left( 
   \theta_1 \cdots \theta_m z^{R_{ij}^{(\ell)}\Lambda}\right) \\ 
&= \frac{1}{f_{\Lambda}} \sum_{1\leq i < j \leq N} 
  \sum_{\ell=0}^{{\cal{S}}_{ij}^{\Lambda} } 
\frac{v_{\Lambda_{ij} \left(R_{ij}^{(\ell)}\Lambda \right)_{ij}}} 
{f_{\left(R_{ij}^{(\ell)} \Lambda\right)_{ij}}} 
\sum_{\sigma \in S_N} {\rm{sgn}}(\sigma_{R_{ij}^{(\ell)}\Lambda}^a) \, 
\mc{K}_{\sigma} \mc{K}_{\sigma_{R_{ij}^{(\ell)}\Lambda}^a} 
\mc{K}_{\sigma_{R_{ij}^{(\ell)}\Lambda}^s}\left( 
   \theta_1 \cdots \theta_m z^{\overline{R_{ij}^{(\ell)}\Lambda}}\right) \\ 
&= \frac{1}{f_{\Lambda}} \sum_{1\leq i < j \leq N} 
  \sum_{\ell=0}^{{\cal{S}}_{ij}^{\Lambda}} 
  {\rm{sgn}}(\sigma_{R_{ij}^{(\ell)}\Lambda}^a) \, 
\frac{v_{\Lambda_{ij} \left(R_{ij}^{(\ell)}\Lambda \right)_{ij}}} 
{f_{\left(R_{ij}^{(\ell)} \Lambda\right)_{ij}}} 
\sum_{\sigma' \in S_N} 
\mc{K}_{\sigma'} \left( 
   \theta_1 \cdots \theta_m z^{\overline{R_{ij}^{(\ell)}\Lambda}}\right) \\ 
&= \frac{1}{f_{\Lambda}} \sum_{1\leq i < j \leq N} 
  \sum_{\ell=0}^{{\cal{S}}_{ij}^{\Lambda}} 
{\rm{sgn}}(\sigma_{R_{ij}^{(\ell)}\Lambda}^a)\, 
\frac{v_{\Lambda_{ij} \left(R_{ij}^{(\ell)}\Lambda \right)_{ij}}} 
{f_{\left(R_{ij}^{(\ell)} \Lambda\right)_{ij}}} 
f_{\overline{R_{ij}^{(\ell)}\Lambda}} \, \, 
m_{\overline{R_{ij}^{(\ell)}\Lambda}} \,. 
\end{split} 
\end{equation} 
Let us be more specific about some of these steps.  In the
fourth equality, we have introduced a factor
$(1+\mc{K}_{ij})/2$ in order to generate two terms out of $\theta_1 \cdots \theta_m
z^{\Lambda}$ because the action of  $B_{ij}+F_{ij}$ is defined on two terms (cf. eqs
\reff{leka}, \reff{lekaa} and \reff{lekaaa}).  We have thus generated an 
intermediate $N=2$
problem in order to use the
$N=2$ results derived previously. The factor $\beta$ is now reabsorbed in the coefficient
$v_{\Lambda_{ij} \left(R_{ij}^{(\ell)}\Lambda \right)_{ij}}$:
\begin{equation}
v_{\Lambda_{ij} \left(R_{ij}^{(\ell)}\Lambda \right)_{ij}}= 2\beta\,\big(
|\Lambda_i-\Lambda_j|-\delta \ell\big)\; , \qquad \ell\not=0\,.
\end{equation}
Observe that the non-diagonal coefficients correspond to the cases where $\ell>0$,
 but that
the action of $B+F$ also includes a diagonal contribution, to be analysed later. In the
sixth equality, we reexpress the outcome in terms of a single  term.
However,
the ordering of the variables may not be adequate for this to be a genuine supermonomial
leading term. Some reordering may be required, forcing the appearance of  
extra factors
$\mc{K}_{\sigma^a}  \mc{K}_{\sigma^s}$ in the next equality. 
Moreover, reordering the anticommuting
variables may generate a sign given by the factor
${\rm{sgn}}(\sigma^a)$. In the following step, we redefine the
summation  variable as follows: ${\sigma'}= {\sigma} {\sigma^a} 
{\sigma^s}$. In the final equality,  we simply  use the
definition of supermonomials. 
 
We are now in a position to compute the coefficient $v_{\Lambda 
\Omega}$. Let $\#(\Omega,\Lambda)$ denote the number of distinct 
ways we can choose $(i,j)$, $i<j$, such that $\Omega = 
\overline{R_{ij}^{(\ell)} \Lambda}$.  Note that such 
``symmetries'' can only occur on the symmetric side, that is, if 
there is more than one possible choice of $i$ (or $j$), then $i>m$ 
(or $j>m$).  Therefore,  
${\rm{sgn}}(\sigma_{R_{ij}^{(\ell)}\Lambda}^a)$ is independent of the particular choice of 
the pair $(i,j)$ leading to $\Omega$, since the underlying degeneracy is independent of
the  antisymmetric side.  The coefficient of $m_{\Omega}$ is thus 
\begin{equation} 
v_{\Lambda \Omega}= 
\frac{1}{f_{\Lambda}} 
{\rm{sgn}}(\sigma_{R_{ij}^{(\ell)}\Lambda}^a) \, 
\frac{v_{\Lambda_{ij} \left(R_{ij}^{(\ell)}\Lambda \right)_{ij}}} 
{f_{\left(R_{ij}^{(\ell)} \Lambda\right)_{ij}}}  
  f_{\Omega} \, \#(\Omega,\Lambda) \, . 
\end{equation} 
The factor $\#(\Omega,\Lambda)$ comes from the different values of the 
pair $(i,j)$ 
leading to the same superpartition $\Omega$.  From Remark~\ref{remark},  
these distinct values all lead to the same coefficient 
$ 
v_{\Lambda_{ij} \left(R_{ij}^{(\ell)}\Lambda \right)_{ij}}/ 
f_{\left(R_{ij}^{(\ell)} \Lambda\right)_{ij}} $. 
 Since it has also just been 
mentioned 
  that these different choices do not affect the value of 
${\rm{sgn}}(\sigma_{R_{ij}^{(\ell)}\Lambda}^a)$, $v_{\Lambda \Omega}$ is simply given by
the coefficient of $m_{\overline{R_{ij}^{(\ell)}\Lambda}}$ in \reff{bigeq} (disregarding
the sum) multiplied by  
$\#(\Omega,\Lambda)$. 
 
Let us now evaluate explicitly the various factors entering in the expression of
$v_{\Lambda \Omega}$, for the three cases treated separately.  We first consider case
III, that is, the case where
$i,j
\in
\{m+1,\dots,N
\}$.  If $\Lambda_i-\ell \neq \Lambda_j+\ell$, when going 
from $\Lambda$ to $\Omega$, the number of $\Lambda_i$ and $\Lambda_j$ 
decreases by 1, whereas the number of $\Lambda_i-\ell$ and 
$\Lambda_j+\ell$  increases by 1 (while the other $\Lambda_k$'s remain unaffected). 
Therefore, the ratio $f_{\Omega}/f_{\Lambda}$ depends only upon the 
multiplicity factors 
involving these parts, that is, 
\begin{equation} 
\begin{split} 
\frac{f_{\Omega}}{f_{\Lambda}}& = 
\frac{ 
n_{\omega^s}(\Lambda_i-\ell)!\, n_{\omega^s}(\Lambda_j+\ell)!\,  
n_{\omega^s}(\Lambda_i)!\, n_{\omega^s}(\Lambda_j)! } 
{n_{\lambda^s}(\Lambda_i-\ell)!\, n_{\lambda^s}(\Lambda_j+\ell)!\,  
n_{\lambda^s}(\Lambda_i)!\, n_{\lambda^s}(\Lambda_j)!} \\ 
& = 
\frac{ 
n_{\omega^s}(\Lambda_i-\ell)!\, n_{\omega^s}(\Lambda_j+\ell)!\,  
n_{\omega^s}(\Lambda_i)!\, n_{\omega^s}(\Lambda_j)! } 
{\bigl( n_{\omega^s}(\Lambda_i-\ell)-1\bigr)!\,\bigl(n_{\omega^s}(\Lambda_j+\ell) 
-1\bigr)! \, 
\bigl(n_{\omega^s}(\Lambda_i)+1\bigr)!\bigl(n_{\omega^s}(\Lambda_j)+1\bigr)!} \\ 
& = \frac{n_{\omega^s}(\Lambda_i-\ell)n_{\omega^s}(\Lambda_j+\ell)} 
{\bigl(n_{\omega^s}(\Lambda_i)+1\bigr)\bigl(n_{\omega^s}(\Lambda_j)+1\bigr) } \, . 
\end{split} 
\end{equation} 
Finally, since 
in that case 
$\#(\Omega,\Lambda)$ is equal to $\bigl(n_{\omega^s}(\Lambda_i)+1\bigr) 
\bigl(n_{\omega^s}(\Lambda_j)+1\bigl)$ (the 
product of the number of $\Lambda_i$'s and $\Lambda_j$'s in 
the symmetric part of $\Lambda$), and  
since $f_{\left(R_{ij}^{(\ell)} \Lambda\right)_{ij}}=1$, we get 
\begin{equation} 
\label{symm3} 
v_{\Lambda \Omega}= 
{\rm{sgn}}(\sigma_{R_{ij}^{(\ell)}\Lambda}^a) \, 
v_{\Lambda_{ij} \left(R_{ij}^{(\ell)}\Lambda \right)_{ij}} 
n_{\omega^s}(\Lambda_i-\ell)n_{\omega^s}(\Lambda_j+\ell) \, , 
\end{equation} 
which agrees with the expression given in Theorem 2.  In the case where $\Lambda_i-\ell =
\Lambda_j+\ell$, when going  from $\Lambda$ to $\Omega$, the number of $\Lambda_i$ and
$\Lambda_j$ 
 decreases by 1, whereas the number of $\Lambda_i-\ell$ 
increases by 2, that is, 
\begin{equation} 
\frac{f_{\Omega}}{f_{\Lambda}}= \frac{n_{\omega^s}(\Lambda_i-\ell) 
\bigl(n_{\omega^s}(\Lambda_i-\ell)-1\bigr)} 
{\bigl(n_{\omega^s}(\Lambda_i)+1\bigr)\bigl(n_{\omega^s}(\Lambda_j)+1\bigr)} \, . 
\end{equation} 
The number $\#(\Omega,\Lambda)$ is still  
$\bigl(n_{\omega^s}(\Lambda_i)+1\bigr)\bigl(n_{\omega^s}(\Lambda_j)+1\bigr)$, 
but now $f_{\left(R_{ij}^{(\ell)} \Lambda\right)_{ij}}=2$, giving 
\begin{equation} 
v_{\Lambda \Omega}= 
\frac{1}{2}{\rm{sgn}}(\sigma_{R_{ij}^{(\ell)}\Lambda}^a) \, 
v_{\Lambda_{ij} \left(R_{ij}^{(\ell)}\Lambda \right)_{ij}} 
n_{\omega^s}(\Lambda_i-\ell)\bigl(n_{\omega^s}(\Lambda_i-\ell)-1 \bigr) \, , 
\end{equation} 
as announced. 
 
Let us now consider case II with $\Lambda_i> \Lambda_j$.  This case 
is similar to case III with 
$\Lambda_i-\ell \neq \Lambda_j+\ell$.  The only difference is that since 
$\Lambda_i$ now belongs to the antisymmetric sector, it does not influence the 
value of $f_{\Omega}/f_{\Lambda}$.  Therefore, in this case, 
\begin{equation} 
\frac{f_{\Omega}}{f_{\Lambda}}= \frac{ 
n_{\omega^s}(\Lambda_j+\ell)} 
{n_{\omega^s}(\Lambda_j)+1} \, , 
\end{equation} 
and 
\begin{equation} 
v_{\Lambda \Omega}= 
{\rm{sgn}}(\sigma_{R_{ij}^{(\ell)}\Lambda}^a) \, 
v_{\Lambda_{ij} \left(R_{ij}^{(\ell)}\Lambda \right)_{ij}} 
n_{\omega^s}(\Lambda_j+\ell)  . 
\end{equation} 
The situation when $\Lambda_i< \Lambda_j$ is 
similar.

Finally, in  case I, $\Lambda_i$ and $\Lambda_j$ both belong to the 
antisymmetric sector, and thus $f_{\Omega}=f_{\Lambda}$.  This gives 
\begin{equation} 
v_{\Lambda \Omega}= 
{\rm{sgn}}(\sigma_{i,j,\ell}^a) \, 
v_{\Lambda_{ij} \left(R_{ij}^{(\ell)}\Lambda \right)_{ij}}  . 
\end{equation}

\subsection{The diagonal coefficients}

To complete the study of the action of the Hamiltonian on the supermonomial basis, we now
evaluate the diagonal coefficient explicitly. The full stCMS Hamiltonian reads
\begin{equation}
\bar {\cal{H}}= A+\beta(B+F)\,,\qquad \text{with}
\qquad A=\sum_{i=1}^N(z_i\partial_i)^2\, .
\end{equation}
The action of $A$ on the supermonomial basis is diagonal:
\begin{equation}
A\,  m_\Lambda=\left(\sum_{i=1}^N\Lambda_i^2 \right) m_\Lambda \, .
\end{equation}
There is also a contribution coming from the
part $\beta(B+F)$ of the Hamiltonian. It can obtained from
 \reff{bigeq} for the case
$\ell=0$ with 
\begin{equation}
v_{\Lambda_{ij} \left(R_{ij}^{(0)}\Lambda \right)_{ij}}= v_{\Lambda_{ij}
\Lambda_{ij}}= \beta|\Lambda_i- \Lambda_j| \, .
\end{equation}
This value is independent of the sector -- cf. eqs \reff{lafi}, \reff{lafia} and
\reff{lafiaa}). The diagonal contribution of $\beta(B+F)m_\Lambda$ is thus
\begin{equation}
\begin{split}
 v_{\Lambda\Lambda}  &=  \frac{1}{f_{\Lambda}} \sum_{1\leq i<j\leq N} 
v_{\Lambda_{ij} \Lambda_{ij}} f_{\Lambda} 
= \beta \sum_{1\leq i<j\leq N} |\Lambda_{i} -\Lambda_{j}| \\ &=  
\beta\sum_{1\leq i<j\leq N}(\Lambda_i^* - \Lambda_j^*) = \beta\sum_{k=1}^N
(N+1-2k)\Lambda_k^*\;.
\end{split}
\end{equation}
The  step in which we eliminate the alsolute value provides  the technical reason
why the  eigenvalue is ultimately expressed in terms of the parts of
$\Lambda^*$. 
The last
equality is obtained as follows \cite{Sutherland:1971ic}. Let $L$ be such that $\Lambda_L^*\not=0$ but
$\Lambda_{j>L}^*=0$. We can write
\begin{equation}
\begin{split}
\sum_{1\leq i<j\leq N}(\Lambda_i^*- \Lambda_j^*) &= 
\sum_{1\leq i\leq L< j\leq N} \Lambda_i^* + \sum_{1\leq i<j\leq L}
(\Lambda_i^*- \Lambda_j^*)\\
&= 
\sum_{1\leq i\leq L} (N-L)\Lambda_i^*  + \sum_{1\leq k\leq L}
[(L-k)-(k-1)] \Lambda_k^* \\
&= 
\sum_{1\leq k\leq L} (N+1-2k)\Lambda_k^* \,.
\end{split}
\end{equation}
To evaluate the
double sum,  we  simply count the number of occurences of $\Lambda_k^*$ with positive and
negative signs: it arises
$N-k$ times with a plus sign and $k-1$ times with a minus sign. Note
 finally that the last
sum can be extended from $L$ to $N$ without changing its value.

Combining the diagonal contribution of $\beta(B+F)$ to that of $A$
yields
\begin{equation}
\varepsilon_\Lambda= \sum_{k=1}^N [ \Lambda_k^2 +  \beta (N+1-2k)\Lambda_k^*]=
\sum_{k=1}^N [ {\Lambda_k^*}^2 +  \beta (N+1-2k)\Lambda_k^*] 
\end{equation}
This is also the eigenvalue of the Jack superpolynomial $\mc{J}_{\Lambda}$,
which is thus computed here without the help of Dunkl operators.

\section{Determinantal expression for the Jack superpolynomials} 
 
\subsection{A new definition of the Jack superpolynomials}
 
Given that we have obtained a partial ordering on superpartitions, 
it is natural to ask whether we can replace Definition~\ref{defjack} 
of the Jack superpolynomials by a {\it {stronger}}  definition 
involving 
the {\it weaker} ordering  on superpartitions. 
That is, we would like to know  whether the Jack 
superpolynomials 
can be characterized by the two conditions: 
\begin{equation} \label{condinew} 
\begin{array}{rl} 
(i): & \quad \mc{J}_{\Lambda} = m_{\Lambda} + \sum_{\Omega;\Omega<^s\Lambda} 
c_{\Lambda 
\Omega} m_{\Omega}\cr (ii): & \quad 
\bar {\cal{H}}\, \mc{J}_{\Lambda}= \varepsilon_{\Lambda} 
J_{\Lambda} \, . 
\end{array} 
\end{equation} 
The interest of such a definition is that, given the triangular action of $\bar 
{\cal{H}}$ (see Corollary~\ref{corotri}) and the fact 
that $\varepsilon_{\Lambda}\neq\varepsilon_{\Omega}$ if $\Omega <^s \Lambda$, 
  we can obtain, as we will see later, 
a determinantal 
expression for $\mc{J}_{\Lambda}$.  Therefore, {\it{there exist functions 
satisfying conditions \reff{condinew}}}. 
That we can indeed characterize the Jack superpolynomials using 
these conditions relies on the uniqueness of the Jack 
superpolynomials (cf. Definition 1), which we now establish. 
 
\begin{lemma} There exists a unique function 
$\mc{J}_{\Lambda}$ 
satisfying the two conditions: 
\begin{equation} 
\begin{array}{rl} 
(i): & \quad \mc{J}_{\Lambda} = m_{\Lambda} +  
\sum_{\Omega; \, \Omega^*< \Lambda^*} c_{\Lambda 
\Omega} m_{\Omega}\cr (ii): & \quad 
\bar {\cal{H}}\, \mc{J}_{\Lambda}= \varepsilon_{\Lambda} 
J_{\Lambda} \, . 
\end{array} 
\end{equation} 
\end{lemma} 
\noindent {\it Proof:}  The existence of such a function will follow from 
the determinantal expression that we will present later on.  The uniqueness is 
proved as follows.  Let ${\mc{J}}_{\Lambda}$ and  
$\bar {\mc{J}}_{\Lambda}$ be two functions satisfying the conditions 
of the theorem.  Then, if we suppose that ${\mc{J}}_{\Lambda}\neq  
\bar {\mc{J}}_{\Lambda}$, we have  
\begin{equation} 
\label{ppetit} 
\mc{J}_{\Lambda} -\bar {\mc{J}}_{\Lambda} =  
\sum_{\Omega; \, \Omega^*< \Lambda^*} 
d_{\Lambda \Omega} m_{\Omega} \, , 
\end{equation} for some coefficients $d_{\Lambda \Omega}$. 
Now, let $\Omega^{(1)}<^T\Omega^{(2)}<^T\dots<^T\Omega^{(n)}$  
be a total ordering 
(compatible with the ordering on superpartitions) of all the superpartitions 
$\Omega^{(i)}$, $i=1,\dots,n$, such that $d_{\Lambda \Omega^{(i)}}\neq 0$. 
Since ${\mc{J}}_{\Lambda}$ and  
$\bar {\mc{J}}_{\Lambda}$ are both eigenfunctions of $\bar{\mc{H}}$ with eigenvalue 
$\varepsilon_{\Lambda}$, we must have $\bar {\cal{H}}\,\left(\mc{J}_{\Lambda} -\bar {\mc{J}}_{\Lambda} \right) = 
\epsilon_{\Lambda} \,\left(\mc{J}_{\Lambda} -\bar {\mc{J}}_{\Lambda} \right) $, 
that is, from Corollary~\ref{corotri}, 
\begin{equation} 
\label{contra} 
\bar {\cal{H}}\, \sum_{i=1}^n 
d_{\Lambda \Omega^{(i)}} m_{\Omega^{(i)}} = \sum_{i=1}^n 
d_{\Lambda \Omega^{(i)}} \left( \varepsilon_{\Omega^{(i)}}\, m_{\Omega^{(i)}}+ 
\sum_{\Gamma <^s \Omega^{(i)}} v_{\Omega^{(i)} \Gamma}m_{\Gamma} \right)  
=\varepsilon_{\Lambda} 
\, \sum_{i=1}^n 
d_{\Lambda \Omega^{(i)}} m_{\Omega^{(i)}} \, . 
\end{equation} 
In the middle expression, the  
coefficient of $m_{\Omega^{(n)}}$ is simply equal to 
$\varepsilon_{\Omega^{(n)}} d_{\Lambda \Omega^{(n)}} $, since  
$\Omega^{(n)}$ dominates $\Omega^{(i)}$, $i=1,\dots,n-1$, in the 
order on superpartitions.  But in the last expression of \reff{contra}, the  
coefficient of $m_{\Omega^{(n)}}$ is equal to 
$\varepsilon_{\Lambda} d_{\Lambda \Omega^{(n)}} $. We must thus have 
$\varepsilon_{\Lambda} d_{\Lambda \Omega^{(n)}}  
=\varepsilon_{\Omega^{(n)}} d_{\Lambda \Omega^{(n)}} $, with  
$d_{\Lambda \Omega^{(n)}}\neq 0$, that is we must have 
$\varepsilon_{\Lambda}   
=\varepsilon_{\Omega^{(n)}} $.  But this is impossible because, 
from \reff{ppetit},  
$\Lambda^* > \left(\Omega^{(n)}\right)^*$, which implies that 
$\varepsilon_{\Lambda}   
\neq\varepsilon_{\Omega^{(n)}} $.  Therefore, a contradiction  arises
and $\mc{J}_{\Lambda}$ cannot be different from $\bar{\mc{J}}_{\Lambda}$. 
 \hfill $\blacksquare$

  It is easy to see, 
from $\reff{condordering}(i)$, that if a function $\mc{J}_{\Lambda}$ 
is such that 
$\mc{J}_{\Lambda}= m_{\Lambda} + \sum_{\Omega<^s\Lambda} c_{\Lambda 
\Omega} m_{\Omega}$, then it is also such that 
$\mc{J}_{\Lambda} 
=m_{\Lambda} + \sum_{\Omega; \, \Omega^*<\Lambda^*} e_{\Lambda 
\Omega} m_{\Omega}$ , 
for some coefficients $e_{\Lambda 
\Omega}$. Therefore, the previous lemma (combined with
the existence of functions satisfying conditions \reff{condinew}) leads directly to the following theorem. 
\begin{theorem} 
The Jack superpolynomials can be defined by the two conditions: 
\begin{equation} 
\begin{array}{rl} 
(i): & \quad \mc{J}_{\Lambda} = m_{\Lambda} + \sum_{\Omega;\Omega<^s\Lambda} 
c_{\Lambda 
\Omega} m_{\Omega}\cr (ii): & \quad 
\bar {\cal{H}}\, \mc{J}_{\Lambda}= \varepsilon_{\Lambda} 
J_{\Lambda} \, . 
\end{array} 
\end{equation} 
\end{theorem} 

\subsection{Determinantal formulas}

As mentioned earlier, an important offshoot of this new definition 
is that we can obtain determinantal expressions for the Jack 
superpolynomials. Indeed, because we are looking for Jack 
superpolynomials with a triangular expansion 
\begin{equation} 
\mc{J}_{\Lambda} = m_{\Lambda} + \sum_{\Omega<^s\Lambda} c_{\Lambda 
\Omega} m_{\Omega} \, , 
\end{equation} 
the fact that $\varepsilon_{\Omega} \neq \varepsilon_{\Lambda}$ if 
$\Omega<^s\Lambda$ 
combined with the triangular action \reff{expatri} 
\begin{equation} 
\bar {\cal{H}} \, m_{\Lambda}= \varepsilon_{\Lambda} 
m_{\Lambda}+\sum_{\Omega <^s \Lambda} v_{\Lambda \Omega} 
\, m_{\Omega} \, , 
\end{equation} 
leads to a determinantal expression. 
 
\begin{theorem} If $\Lambda^{(1)},\Lambda^{(2)},\ldots, 
\Lambda^{(n)}=\Lambda$ is 
a total ordering (compatible with the ordering on superpartitions) 
of all superpartitions $\leq^s\Lambda$, then 
the Jack superpolynomial $\mc{J}_{\Lambda}$ is given by the 
following determinant: 
\begin{equation} 
\mc{J}_{\Lambda} = {c_{\Lambda}} 
\left| \begin{array}{cccccc} 
m_{\Lambda^{(1)}} & m_{\Lambda^{(2)}} &\cdots & \cdots 
& m_{\Lambda^{(n-1)}} & m_{\Lambda^{(n)}} \cr 
\varepsilon_{\Lambda^{(1)}}-\varepsilon_{\Lambda^{(n)}} 
& v_{\Lambda^{(2)}\Lambda^{(1)}}&\cdots & 
\cdots & v_{\Lambda^{(n-1)}\Lambda^{(1)}} & v_{\Lambda^{(n)}\Lambda^{(1)}} \cr 
                                   0& 
\varepsilon_{\Lambda^{(2)}}-\varepsilon_{\Lambda^{(n)}}& \cdots& \cdots & 
v_{\Lambda^{(n-1)}\Lambda^{(2)}} 
  & v_{\Lambda^{(n)}\Lambda^{(2)}} \cr 
                 \vdots &0 &\ddots &&\vdots& \vdots \cr 
                 \vdots &\vdots &\ddots &\ddots& & \vdots \cr 
0 & 0 &\cdots & 0 & 
\varepsilon_{\Lambda^{(n-1)}}-\varepsilon_{\Lambda^{(n)}}  & 
v_{\Lambda^{(n)}\Lambda^{(n-1)}} 
\end{array} \right| \, , 
\end{equation} 
where the constant of proportionality is 
\begin{equation} 
c_{\Lambda} = (-1)^{n-1} \prod_{i=1}^{n-1} 
\frac{1}{\varepsilon_{\Lambda^{(i)}}-\varepsilon_{\Lambda^{(n)}}} \, . 
\end{equation} 
\end{theorem} 
\noindent For a proof that this determinant is in fact a non-zero 
eigenvector of $\bar {\cal{H}} $ with eigenvalue 
$\varepsilon_{\Lambda}$, the reader is referred to 
\cite{LapointeLascouxMorse}.  The proof found in this article can 
be applied to our case as well. 
Notice that in the determinantal expression, the relative ordering 
of the superpartitions not related by the partial 
ordering $\geq^s$ is irrelevant. 
 
   It might be important to remark that, since the coefficients 
$v_{\Lambda \Omega}$ do not depend on the number of variables $N$, and since 
the differences 
\begin{equation} 
\varepsilon_{\Omega}-\varepsilon_{\Lambda}= 
\varepsilon_{\Omega^*}-\varepsilon_{\Lambda^*}= 
  \sum_{k} 
  \left( 
  ({\Omega^*_k}^2-{\Lambda^*_k}^2) - 2 \, k \,
\beta\,(\Omega^*_k-\Lambda^*_k) \right) \, , 
\qquad |\Omega|=|\Lambda| \, , 
\end{equation} 
also do not depend on the number of variables, the 
determinantal expression for $\mc{J}_{\Lambda}$ does not depend on 
$N$. 
 
For example, with 
\begin{equation} 
\begin{split} 
\bar {\cal{H}}\, m_{(3;1)}&= (10+4 \, \beta \,  N -6\, \beta) \, m_{(3;1)} + 
  2 \, \beta\, m_{(2;2)} 
+ 8 \, \beta \, m_{(2;1,1)}+2 \, \beta \, m_{(1;2,1)} \\ 
\bar {\cal{H}} \, m_{(2;2)}&= (8+4\, \beta \, N -8\, \beta) \, m_{(2;2)} + 
  4 \, \beta \, m_{(2;1,1)}+2 \, \beta \, m_{(1;2,1)} \\ 
\bar {\cal{H}} \, m_{(2;1,1)}&= (6+4\, \beta \, N -10\, \beta) \, m_{(2;1,1)} + 
6 \, \beta \, m_{(1;1,1,1)} \\ 
\bar {\cal{H}}\,  m_{(1;2,1)}&= (6+4\, \beta \, N -10\, \beta) \, 
m_{(1;2,1)} + 
12 \, \beta \, m_{(1;1,1,1)} \\ 
\bar {\cal{H}}\,  m_{(1;1,1,1)}&= (4+4\, \beta\,  N -16\, \beta) \, 
m_{(1;1,1,1)} \, , 
\end{split} 
\end{equation} 
we have 
\begin{equation} 
\begin{split} 
\mc{J}_{(3;1)} & = \frac{1}{64(3+5\beta)(1+\beta)^3} 
\left| \begin{array}{ccccc} 
  m_{(1;1,1,1)} & m_{(1;2,1)}& m_{(2;1,1)} & m_{(2;2)} & m_{(3;1)} \cr 
-6-10 \beta & 12 \beta & 6 \beta & 0 & 0 \cr 
0 & -4 -4 \beta & 0 & 2 \beta & 2 \beta \cr 
0 & 0 & -4 -4\beta & 4 \beta & 8 \beta \cr 
0 & 0 & 0 & -2 -2\beta & 2 \beta 
\end{array}\right| \\ 
& = m_{(3;1)}+\frac{\beta}{1+\beta} m_{(2;2)}+ 
\frac{\beta(2+3\beta)}{(1+\beta)^2} m_{(2;1,1)}+ 
\frac{\beta(1+2\beta)}{2(1+\beta)^2} m_{(1;2,1)}+ 
\frac{3\beta^2}{(1+\beta)^2} m_{(1;1,1,1)} \, . 
\end{split} 
\end{equation}

\section{Conclusion} 
 
Although unravelling the natural generalization of the dominance ordering at 
the level of superpartitions was our original motivation for this 
work, we have obtained much more than that: 
  we ended up with a simple determinantal expression for the Jack 
superpolynomials. In a sense, the mere discovery of a determinantal 
expression for Jack superpolynomials ensures that the pivotal 
concepts underlying their construction, 
namely monomial symmetric superpolynomials and 
superpartitions (as well as their associated ordering) are 
the right ones. 

There is a natural supersymmetric extension of the physical scalar product with
respect to which the Jack polynomials are orthogonal.   However, the Jack superpolynomials 
are not orthogonal with respect to this supersymmetric scalar product.  
The next step is thus to seek a general pattern for 
constructing linear combinations of Jack superpolynomials sharing
the same Hamiltonian eigenvalue 
that would be orthogonal.

Also, we believe the present formalism to be extendable to the
case of the
Hi-Jack polynomials (or generalized Hermite polynomials) 
and their supersymmetric counterparts. We hope to report
elsewhere on this subject.

\vskip0.3cm 
\noindent {\bf NOTE ADDED}

After this article got accepted for publication, we completed the program of constructing
{\it orthogonal} Jack superpolynomials.  These results will be presented in the article 
{\it Jack polynomials in superspace}.

\vskip0.3cm 
\noindent {\bf ACKNOWLEDGMENTS}

This work was  supported by  FCAR 
and NSERC. L.L. wishes to thank Luc Vinet for his 
financial support. P.D. would also like to thank the Fondation 
J.A Vincent 
for a student 
fellowship. 
 
\appendix 
 
\section{Proofs of Identity~\ref{iden1} and~\ref{iden2}} 
 
\noindent {\it{Proof of Identity~\ref{iden1}:}} \quad 
Expanding the factors $z_{12}$ and $z^{12}$, Identity~\ref{iden1} is equivalent 
to 
\begin{equation} 
\begin{split} 
& (r-s) \left(z_1^{r-s+1}-z_1^{r-s}z_2 \right) + 
\sum_{\ell=1}^{r-s-1}  2(r-s-\ell)  z_1^{r-s-\ell+1} z_2^{\ell} 
-\sum_{\ell=1}^{r-s-1} 2(r-s-\ell)  z_1^{r-s-\ell} z_2^{\ell+1} \\ 
& \qquad \qquad \qquad = (r-s) \left(z_1^{r-s+1}+z_1^{r-s}z_2 \right)  -2 
\sum_{\ell=1}^{r-s}  z_1^{r-s-\ell+1} z_2^{\ell} \, . 
\end{split} 
\label{iden11} 
\end{equation} 
Now, doing simple transformations on the indices of summation, we have
\begin{equation} 
\begin{split} 
& (r-s) \left(z_1^{r-s+1}-z_1^{r-s}z_2 \right) + 
\sum_{\ell=1}^{r-s-1}  2(r-s-\ell)  z_1^{r-s-\ell+1} z_2^{\ell} 
-\sum_{\ell=1}^{r-s-1} 2(r-s-\ell)  z_1^{r-s-\ell} z_2^{\ell+1} \\ 
&= 
(r-s) \left(z_1^{r-s+1}-z_1^{r-s}z_2 \right) +\sum_{\ell=1}^{r-s-1} 
2(r-s-\ell)  z_1^{r-s-\ell+1} z_2^{\ell} 
-\sum_{\ell=2}^{r-s} 2(r-s-\ell+1)  z_1^{r-s-\ell+1} z_2^{\ell} \\ 
&= (r-s) \left(z_1^{r-s+1}-z_1^{r-s}z_2 \right) + 2(r-s-1)z_1^{r-s}z_2-2 
z_1 z_2^{r-s} -2 
\sum_{\ell=2}^{r-s-1}  z_1^{r-s-\ell+1} z_2^{\ell} \\ 
& = (r-s) \left(z_1^{r-s+1}+z_1^{r-s}z_2 \right)  -2 
\sum_{\ell=1}^{r-s}  z_1^{r-s-\ell+1} z_2^{\ell} \, , 
\end{split} 
\end{equation} 
which proves \reff{iden11} and, consequently, Identity~\ref{iden1}. 
\hfill $\blacksquare$

\noindent {\it{Proof of Identity~\ref{iden2}:}} \quad  Using 
Identity~\ref{iden1} twice (once with $z_1 \leftrightarrow z_2$), 
we have 
\begin{equation} 
\begin{split} 
& (r-s)z^{12} \left(z_1^{r-s}+z_2^{r-s} \right) - 
4 \sum_{\ell=1}^{r-s}  z_1^{r-s-\ell+1} z_2^{\ell} \\ 
& \quad = z_{12} \left\{  (r-s) \left(z_1^{r-s}-z_2^{r-s} \right) 
+\sum_{\ell=1}^{r-s-1} 2(r-s-\ell)  z_1^{r-s-\ell} z_2^{\ell} 
-\sum_{\ell=1}^{r-s-1} 2(r-s-\ell)  z_1^{\ell} z_2^{r-s-\ell} 
  \right\} \, . 
\end{split} 
\end{equation} 
Finally, after a sequence of simple transformations, 
\begin{equation} 
\begin{split} 
  \sum_{\ell=1}^{r-s-1} & 2(r-s-\ell)  z_1^{r-s-\ell} z_2^{\ell} 
-\sum_{\ell=1}^{r-s-1} 2(r-s-\ell)  z_1^{\ell} z_2^{r-s-\ell} \\ 
& \qquad = \sum_{\ell=1}^{r-s-1} 2(r-s-\ell)  z_1^{r-s-\ell} z_2^{\ell} 
-\sum_{\ell=1}^{r-s-1} 2 \, \ell \, z_1^{r-s-\ell} z_2^{\ell} \\ 
& \qquad = \sum_{\ell=1}^{r-s-1} 2(r-s-2\ell)  z_1^{r-s-\ell} z_2^{\ell} \\ 
& \qquad = \sum_{\ell=1}^{\lfloor \frac{r-s-1}{2} \rfloor} 2(r-s-2\ell) 
z_1^{r-s-\ell} z_2^{\ell} + \sum_{\ell=\lfloor \frac{r-s-1}{2} 
\rfloor+1}^{r-s-1} 2(r-s-2\ell)  z_1^{r-s-\ell} z_2^{\ell} \\ 
& \qquad = \sum_{\ell=1}^{\lfloor \frac{r-s-1}{2} \rfloor} 2(r-s-2\ell) 
z_1^{r-s-\ell} z_2^{\ell} - 
\sum^{r-s-1-\lfloor \frac{r-s-1}{2} \rfloor}_{\ell=1} 2(r-s-2\ell) 
  z_1^{\ell} z_2^{r-s-\ell} \\ 
& \qquad = \sum_{\ell=1}^{\lfloor \frac{r-s-1}{2} \rfloor} 2(r-s-2\ell) 
  \left(z_1^{r-s-\ell} z_2^{\ell} - z_1^{\ell} z_2^{r-s-\ell} \right) \, , 
\end{split} 
\end{equation} 
we obtain the Identity 8. Note that in the final step, 
when $r-s$ is even, we have 
$r-s-1-\lfloor \frac{r-s-1}{2} \rfloor=\lfloor \frac{r-s-1}{2} \rfloor +1$. 
The extra case corresponding to 
$\ell =\lfloor \frac{r-s-1}{2} \rfloor +1= (r-s)/2$ can be ignored since it 
  has a  factor $(r-s-2\ell)=0$. 
\hfill $\blacksquare$


\begin{thebibliography}{99} 
\addcontentsline{toc}{section}{References} 
 
 
 
\bibitem{Calogero:1969ie} 
F.~Calogero, 
\emph{Ground state of one-dimensional N body system}, 
J.\ Math.\ Phys.\  {\bf 10}, 2197 (1969); \emph{Solution of a three body 
problem in one-dimension}, 
J.\ Math.\ Phys.\  {\bf 10}, 2191 (1969) ; \emph{Solution of the 
one-dimensional N body problems with quadratic and/or inversely quadratic 
pair potentials}, 
J.\ Math.\ Phys.\  {\bf 12}, 419 (1971). 
 
\bibitem{Moser1974} 
J.~Moser, \emph{Three integrable {H}amiltonian systems connected with 
              isospectral deformations}, Surveys in applied mathematics, 
235--258 (1976). 
 
\bibitem{Sutherland:1971ic} 
B.~Sutherland, 
\emph{Quantum many body problem in one-dimension: ground state}, 
J.\ Math.\ Phys.\  {\bf 12}, 246 (1971); 
\emph{Exact results for a quantum many body problem in one-dimension}, 
Phys.\ Rev.\  {\bf A4}, 2019 (1971); 
\emph{Exact results for a quantum many body problem in one-dimension. II}, 
Phys.\ Rev.\  {\bf A5}, 1372 (1972). 
 
 
\bibitem{Olshanetsky:1981dk} 
M.~A.~Olshanetsky and A.~M.~Perelomov, 
\emph{Classical integrable finite dimensional systems related to Lie algebras}, 
Phys.\ Rept.\  {\bf 71}, 313 (1981) ; 
\emph{Quantum integrable systems related to Lie algebras}, 
Phys.\ Rept.\  {\bf 94} (1983) 313. 
 
 
\bibitem{Jack1970} 
     {H. Jack}, 
      \emph{A class of symmetric polynomials with a parameter}, 
  {Proc. Roy. Soc. Edinburgh Sect. A}, 
    {\bf69}, 
         {1--18} ({1970/1971}); 
\emph{A surface integral and symmetric functions}, 
{Proc. Roy. Soc. Edinburgh Sect. A}, 
  {\bf69},  {part 4}, {347--364} ({1972}). 
 
 
 
\bibitem{Stanley1988} 
     R. P. Stanley, 
     \emph{Some combinatorial properties of Jack symmetric functions}, Adv. 
Math.,{\bf77}, 76-115 (1988). 
 
 
\bibitem{DLM} 
P.~Desrosiers, L.~Lapointe and P.~Mathieu, \emph{Supersymmetric 
Calogero-Moser-Sutherland models and Jack superpolynomials}, Nucl. 
Phys. {\bf B 606}, 547 (2001). 
 
 

 
 
\bibitem{SriramShastry:1993cz} 
B.~Sriram Shastry and B.~Sutherland, 
\emph{Superlax pairs and infinite symmetries in the  $ 1/r^2 $  system}, 
Phys.\ Rev.\ Lett.\  {\bf 70}, 4029 (1993), cond-mat/9212029. 
 
 
\bibitem{Macdonald1995} 
     {I.~G.~ Macdonald}, 
      \emph{Symmetric functions and {H}all polynomials}, 
2ieme edition, The Clarendon Press Oxford University Press, 
      ({1995}); 
      \emph{Symmetric functions and orthogonal polynomials}, 
      {Dean Jacqueline B. Lewis Memorial Lectures presented at Rutgers 
              University}, 
{American Mathematical Society}, ({1998}). 
 
 
 
 
\bibitem{Sogo} 
K.~Sogo, \emph{Eigenstates of Calogero-Sutherland-Moser model and 
generalized Schur 
functions}, J. Math. Phys. {\bf 35}, 2282 (1994). 
 
 
 
 
\bibitem{LapointeLascouxMorse} 
L.~Lapointe, A.~Lascoux and J.~Morse, \emph{Determinantal formula and recursion 
for Jack polynomials}, Electro. J. Comb. {\bf 7} 467 (2000). 
 
 
 
\bibitem{Polychronakos:1992zk} 
A.~P.~Polychronakos, 
\emph{Exchange operator formalism for integrable systems of particles}, 
Phys.\ Rev.\ Lett.\  {\bf 69}, 703 (1992), hep-th/9202057; 
J.~A.~Minahan and A.~P.~Polychronakos, 
\emph{Integrable systems for particles with internal degrees of freedom}, 
Phys. Lett. {\bf B302} 299 (1993), hep-th/9206046. 
 
\end{thebibliography}
\end{document}